# Tailoring the photovoltaic effect in (111) oriented BiFeO$_3$/LaFeO$_3$ superlattices

J. Belhadi[1], S. Yousfi[1], M. El Marssi[1], D. C. Arnold[2], H. Bouyanfif[1,*]

[1]LPMC EA2081, Université de Picardie Jules Verne, 33 Rue Saint Leu, 80000 Amiens, France

[2]School of Physical Sciences, University of Kent, Canterbury, Kent, CT2 7NH, UK

*corresponding author: houssny.bouyanfif@u-picardie.fr

**Abstract**

Ferroelectric and photovoltaic properties of (BiFeO$_3$)$_{(1-x)\Lambda}$/(LaFeO$_3$)$_{x\Lambda}$ superlattices grown by pulsed laser deposition have been investigated ($\Lambda$ being the bilayer thickness). For a high concentration of BiFeO$_3$ a ferroelectric state is observed simultaneously with a switchable photovoltaic response. In contrast for certain concentration of LaFeO$_3$ a non-switchable photovoltaic effect is evidenced. Such modulation of the PV response in the superlattices is attributed to the ferroelectric to paraelectric phase transition which is controlled with the increase of $x$. Remarkably, concomitant to this change of PV mechanism, a change of the conduction mechanism also seems to take place from a bulk-limited to an interface-limited transport as $x$ increases.

**Introduction**

The multiferroic, BiFeO$_3$ (BFO) has been massively investigated due to the room temperature coexistence of antiferromagnetic and ferroelectric orders[1–3]. BFO presents a rhombohedral *R*3*c* symmetry and adopts a perovskite structure with a pseudo-cubic cell parameter of 3.96Å[2]. Ferroelectricity and antiferromagnetism emerge respectively below T$_C$=1100K and $T_N$=640K. Understanding of the magnetoelectric coupling was motivated by the potential application in information storage technologies[1,3]. A very large number of experimental and theoretical reports have been devoted to thin films and showed the versatile character of BFO under strain (possibility to tune the phases under epitaxial strain) [4–6]. BFO also triggered a rebirth of interest in the photovoltaic (PV) properties of ferroelectric oxides after the observation of above band-gap photo-voltage (open circuit voltage $V_{oc}$>16V) [7,8]. The very large photo-voltage was attributed to a bulk-like non-centrosymmetric mechanism and since then numerous investigations have tried to disentangle the many possible PV mechanisms in thin films (Schottky barriers, defects, ferroelectric domain structure) [9–15]. However, the main drawback of BFOis the low generated photocurrent due to the insulating character of FE materials leading to low conversion efficiency. Recently, by using chemical ordering of $Bi_2FeCrO_6$ thin films Nechache*et al.* demonstrated high power conversion efficiencies up to 8%, whichshows the possibility of employing multiferroic materials in PV devices [16].Maximum PV responses were also evidenced at morphotropic phase boundaries of $(Bi,La)FeO_3$ solid solutions highlighting the extreme sensitivity of the electronic band structure to chemical doping and polar order [17]. The engineering of PV properties in ferroic perovskite oxide superlattices (SLs) can also be considered as one of the promising alternative routes for improving the PV activity in such materials. Modulation and optimization of PV response in such SLs platforms is very likely due to the large number of degrees of freedom present in these systems (e.g. periodicity, nature and ratio of constituents of the bilayers, number of bilayers) and the extreme sensitivity of BFO to external perturbations (such as bandgap engineering and modulation of the structural/ferroelectric properties). Here, we explore the PV activity in epitaxial multiferroic $(BiFeO_3)_{(1-x)\Lambda}/(LaFeO_3)_{x\Lambda}$ $((BFO)_{(1-x)\Lambda}/(LFO)_{x\Lambda})$ SLs grown using the PLD technique on (111) oriented $SrTiO_3$ substrates. An obvious reason for the choice of such crystallographic orientation is the large 100μC/cm$^2$ bulk BFO polarization along the [111] pseudo-cubic direction [14,18]. We focus in this report on the effect of the variation of BFO and LFO layer thickness for a fixed number of bilayers (25) and constant period of approximately Λ=10nm.

Room temperature X-ray diffraction and Raman spectroscopy investigations have previously evidenced a structural change at about $x$=0.5 from a rhombohedral/monoclinic structure for rich BFO to an orthorhombic symmetry for rich LFO systems[19].

**Methods**

Details of the growth and room temperature structure are presented elsewhere and this work focuses solely on the ferroelectric and photovoltaic responses of the SLs[14,19]. Bottom SrRuO$_3$ (30nm) and top ITO electrodes (0.1mm diameters) were used for electrical characterization. The ferroelectric P-E loops measurements were investigated at 1 kHz using a TF Analyzer 1000 aixACCT. The I(V) curves were collected using a Keithley 2635 electrometer. An Argon-Krypton tuneable laser was used to illuminate the samples for the PV measurement (488 nm to 647 nm). The temperature was controlled using a Linkam stage that allows a temperature stability of ±0.1K.

**Results**

Figure 1 presents the ferroelectric hysteresis loops for the $x$=0.1 and 0.3 concentrations of LFO in the bilayers.

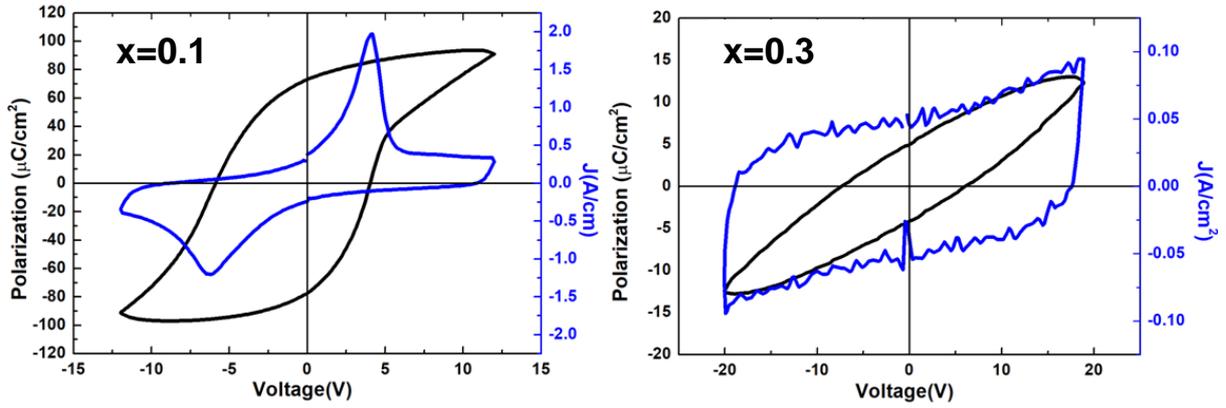

Figure 1. Hysteresis loops for $x=0.1$ and $x=0.3$ $(BFO)_{(1-x)\Lambda}/(LFO)_{x\Lambda}$ SLs measured at room temperature.

The hysteresis loop for $x=0.1$ presents robust ferroelectric character with a remnant polarization of about $Pr=80\mu C/cm^2$ and an imprint toward negative voltages. Imprints may arise from inhomogeneous distribution of vacancies, strain and the different nature of the bottom and top electrodes[10,14]. The $x=0.3$ SLs presents in contrast an extremely weak/no ferroelectric character. These results are in perfect agreement with the published structural (XRD and Raman) investigation showing a change from arhombohedral-like structure to an orthorhombic (*Pnma*-like) structurewith increasing $x$[19].The I(V) characteristics measured in the dark also show a dependence with $x$ as observed in Figure 2.

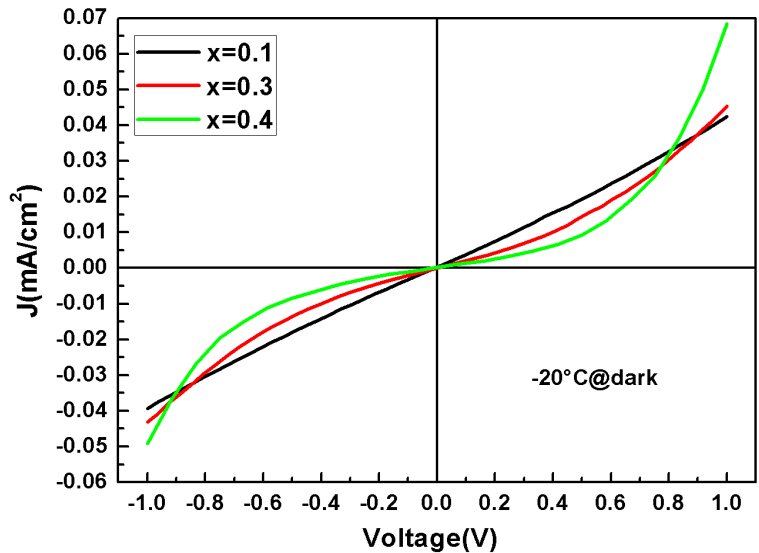

Figure 2. I(V) characteristics for three $(BFO)_{(1-x)\Lambda}/(LFO)_{x\Lambda}$ SLs

The I(V) characteristics present a trend from a linear to a non-linear character with increasing $x$. While the I(V) curve for $x=0.1$ is purely symmetric, the $x=0.4$ is non symmetric with a current flowing preferentially for a positive voltage. Symmetric and non-symmetric I(V) curves are signatures of bulk-limited and interface-limited transport of charges respectively. Figure 2 demonstrates the tailoring of charge transport with the change in $x$ and the polar nature of the SLs. It is important to make a dichotomy between transport in the dark and PV mechanismsas such observations help provide insights into the structure of the electric field within the electrode/multilayer/electrode stacking. Indeed any internal electric field may contribute to the splitting of the photo-induced electron hole (e-h) pairs. In order to evidence any signature of PV activity similar I(V) curves were collected under illumination (514nm wavelength) and are presented in Figure 3.

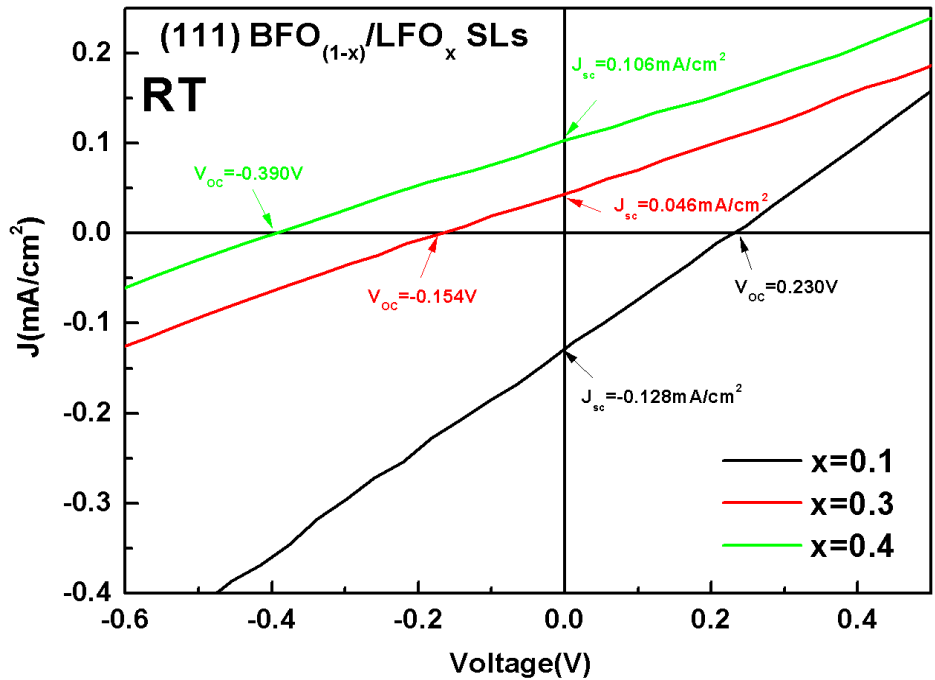

Figure 3. Room temperature I(V) curves collected under 1mW excitation (514nm).

A PV response is without doubt clearly observed in the three un-poled SLs investigated. Non-zero open circuit voltage $V_{oc}$ and short-circuit current $J_{sc}$ are indeed detected. We recall that in the dark the I(V) curves cross the origin (i.e I=0 A when V=0V). The un-poled (or as deposited) PV responses show different signs ($V_{oc}> 0$ for $x$=0.1 and $V_{oc}<0$ for $x$=0.3 and $x$=0.4). This observation is most likely correlated to the different polar (FE vs PE) and transport behaviour (bulk vs interface limited) revealed above for the three SLs. A different PV mechanism is therefore inferred for $x$=0.1 compared to $x$=0.3 and $x$=0.4. The influence of laser power has also been explored and is presented in figure 4.

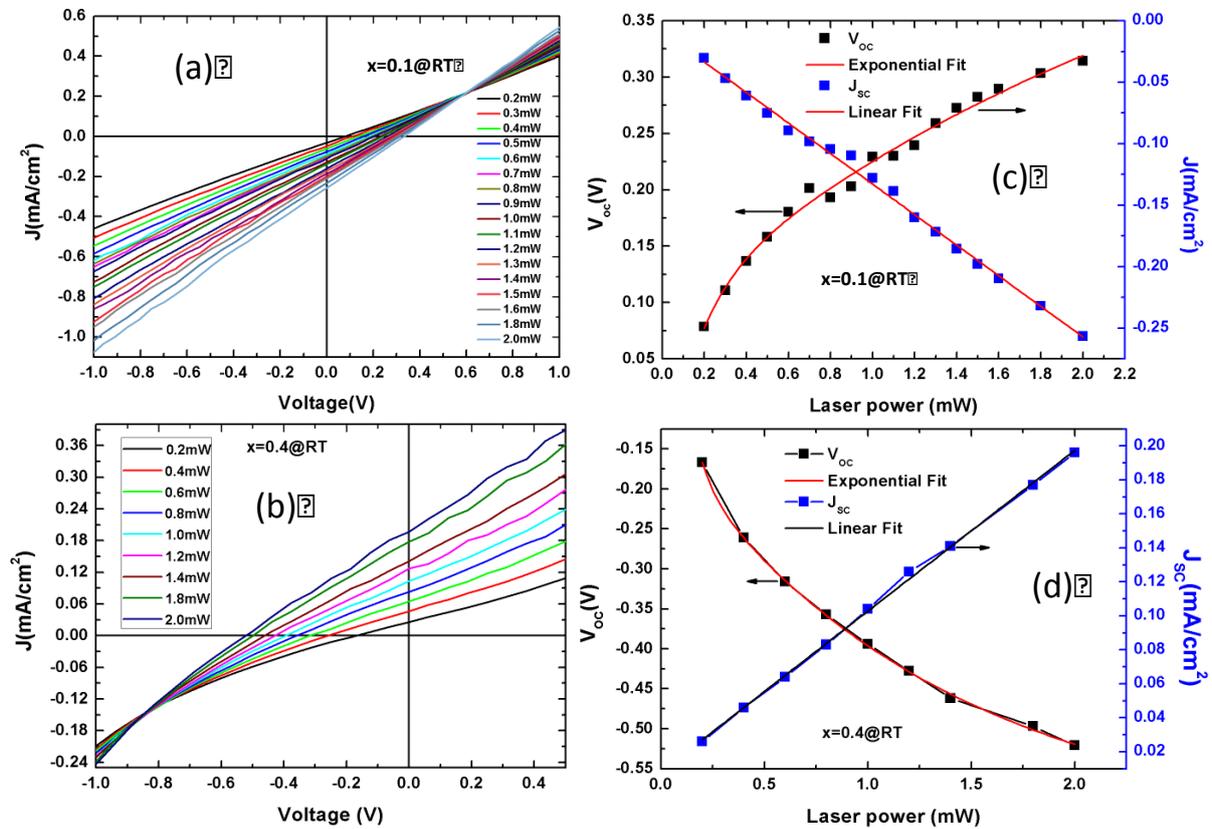

Figure 4. Room temperature influence of laser power on the I(V) characteristics of (a) $x$=0.1 and (b)$x$=0.4 SLs. Corresponding $V_{oc}$ and $J_{sc}$ are shown on (c) and (d).

An increase of $V_{oc}$ and $J_{sc}$ is observed with increasing laser power. The estimated values of $V_{oc}$ and $J_{sc}$ for each SLs are shown in figure 4 (c) and (d) and a saturation of $V_{oc}$ is evidenced for both samples. Such saturation may be explained by the higher probability of e-h recombination above a certain density of photo-induced e-h pairs. Overheating, however, cannot be excluded. $J_{sc}$ on the other hand increases linearly with power for both $x=0.1$ and $x=04$ signalling a direct proportionality between the density of photo-induced e-h pairs with laser power whatever the PV mechanism (*a priori* different for these two SLs) [20]. To better understand the origin of the PV mechanism the influence of temperature has additionally been probed. Figure 5 shows the evolution with temperature of the PV response for $x=0.1$ and $x=04$ SLs.

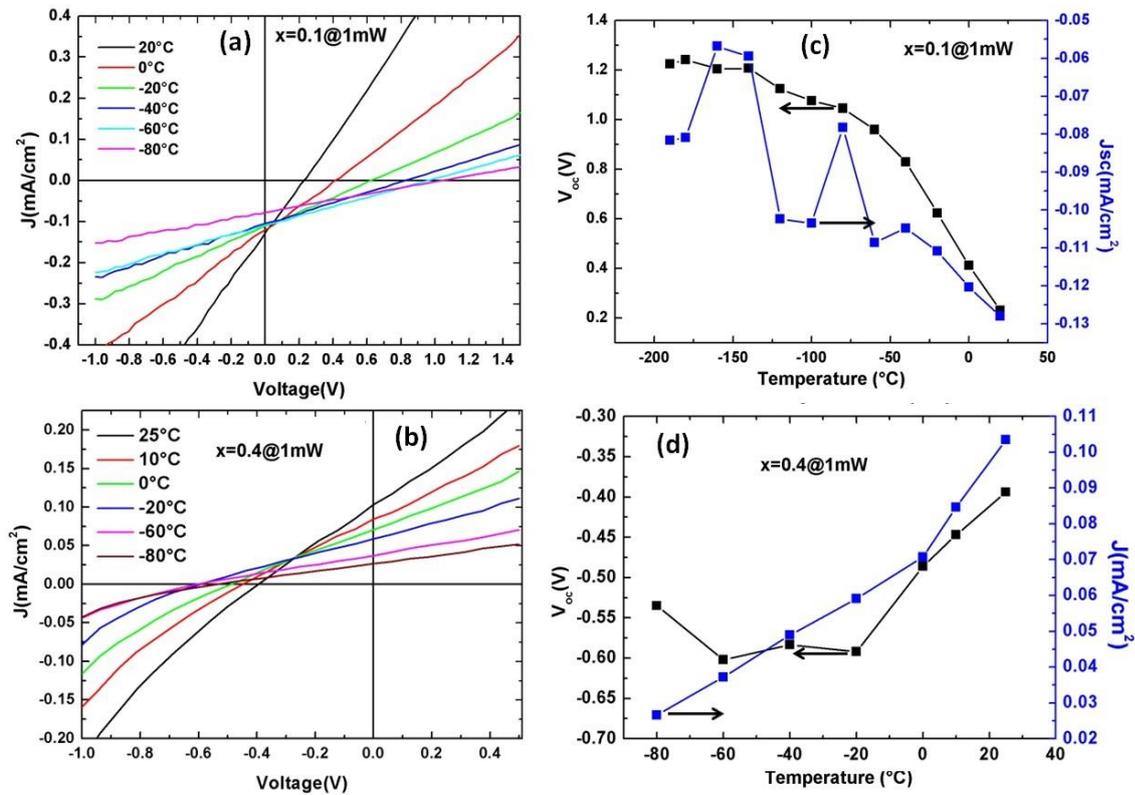

Figure 5. Evolution with temperature of the PV response for (a) $x=0.1$ and (b) $x=0.4$. Corresponding $V_{oc}$ and $J_{sc}$ are shown in (c) and (d) respectively. The PV responses present a change with temperature in particular for the ferroelectric $x=0.1$ SL for which a significant increase of $V_{oc}$ is evidenced. Indeed $V_{oc}$ is about 0.2V at room temperature and increases up to 1.2V at -80°C. In contrast, a limited change is observed for the paraelectric $x=0.4$ SL ($V_{oc}$ varies in the range -0.4V to -0.6V). This is again a hint of a different mechanism behind

the PV response. The increase of $V_{oc}$ on cooling might be attributed to the decrease of leakage currents and the concomitant increase of the inner electric field responsible of the e-h splitting. This increase of the inner field on cooling may be due to the increase of the spontaneous polarization on cooling (hysteresis loops versus temperature will be the subject of another article). The decrease of dark conductivity is indeed often accompanied with an increase of open $V_{oc}$. The observed decrease of $J_{sc}$ on cooling for both SLs is most likely explainedby the increase of the band-gap and the subsequent decrease of the photo-induced e-h pair. This point needs to be confirmed by complementary investigations. To confirm whether the PV response is connected to the ferroelectric polarization, PV measurements were performed after application of voltage pulses (1ms duration) of different signs and amplitude. The application of such pulses prior to the I(V) measurements under illumination enables the preparation of the ferroelectric domain state. The I(V) curves collected after applied pulses and under illumination are presented in Figure 6 for the SLs with *x*=0.1 and *x*=0.4. We recall that *x*=0.1 and *x*=0.4 are respectively ferroelectric and paraelectric SLs (P-E loops and previous structural determination in ref.19).

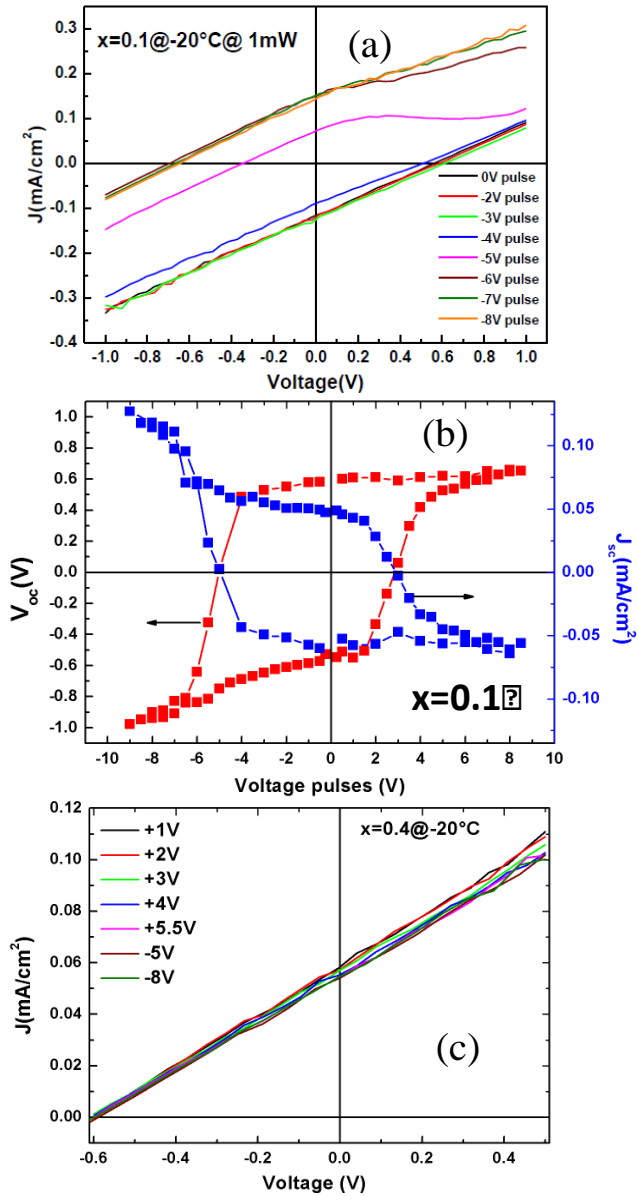

Figure 6. (a) I(V) characteristics under illumination (514nm and 1mW) after application of voltage pulses for *x*=0.1 (b) $V_{oc}$ and $J_{sc}$ versus pre-applied voltage pulses for *x*=0.1 (c) I(V) characteristics under illumination (514nm and 1mW) after application of voltage pulses for *x*=0.4.

While an influence is observed for *x*=0.1 no dependence of the I(V) curves are detected after applications of voltage pulses for *x*=0.4. The I(V) curves for *x*=0.4 remain unchanged and the PV quantities $J_{sc}$ and $V_{oc}$ are unaffected by the pulses. The *x*=0.1 SL demonstrates a switchable PV response with a change in sign when voltage pulses of enough amplitude are used. Detection of a weak non-linearity of the I(V) data after a pulse of -5V is evidenced and may be explained by a back-switching of very mobile ferroelectric domains. Note that positive voltage pulses do no modify at all the I(V) curves for the $x$=0.1 when performed in the as-deposited state (up polarization). However negative pulses of increasing amplitude continuously shift the I(V) characteristics toward the left of I(V) quadrant. Subsequent positive pulses of increasing amplitude shift the I(V) curve back toward the initial position (as measured in the virgin state). Combining the obtained data a hysteresis of Voc and Jsc is measured as shown in Figure 6 (b). The $J_{sc}$ and $V_{oc}$ hysteresis are reminiscent of the ferroelectric hysteresis shown in Figure 1 with a similar shift toward negative values. The imprint observed of $J_{sc}$, $V_{oc}$ and the polarization loop strongly suggests a ferroelectric origin of the PV effect measured for the $x$=01 SL in contrast with the $x$=0.4 SL. Last but not least the coercive voltages for $x$=0.1 (see hysteresis loop Fig. 1) are close to the pulse amplitudes necessary to switch the sign of $V_{oc}$ and $J_{sc}$. PV devices built on a Schottky junction are non-switchable and present asymmetric and non-linear I(V) characteristics and contribution of such Schottky barriers probably explains the photo-induced properties for the $x$=0.4 sample. Figure 6 clearly demonstrates the tailoring of the PV properties in the BFO-LFO based superlattices.

Our work,therefore, providesan efficient way for engineering the photovoltaic effect in ferroelectric materials via the design of new epitaxial superlatticesfor next generation photovoltaic materials.In built field due to top and bottom different electrodes may contribute to the e-h splitting and PV activity[21]. Non-switchable contribution probably arises from this in built field and additional investigations are required to disentangle the different possible mechanisms at the origin of PV behaviour in such artificial complex oxide SLs.

**Conclusions**

In conclusion ferroelectric and photovoltaic responses have been probed in (111) oriented epitaxial $(BFO)_{(1-x)\Lambda}/(LFO)_{x\Lambda}$ SLs. Simultaneously to a *x* induced ferroelectric to paraelectric room temperature transition, a modification of the transport mechanism is suggested. A PV response is demonstrated in all SLs whose origin depends on the polar stability. Ferroelectric SLs show a switchable PV response while paraelectric-like SLs present PV behaviour reminiscent of an Schottky Junction. Emergence of PV responses in such SL platforms clearly calls for further investigation and enables opportunities for devices with giant performances considering the large number of degrees of freedom (number of bilayers, nature of the constituents, relative thicknesses) present in these systems.

**Acknowledgments**

JB and MEM gratefully acknowledges the European commission for funding ENGIMA Project No. 778072. SY is grateful to the French Ministry of Higher Education for the PhD funding. HB is grateful to the Region of Picardy for funding this work through the ZOOM project.